\documentclass[preprint,showpacs,amsmath,amssymb,prd]{revtex4}
\usepackage{dcolumn}
\usepackage{epsfig}
\usepackage{bm}
\begin{document}


\title{\large\bf 
A Novel Scheme to Search for Fractional Charge Particles
in Low Energy Accelerator Experiments 
} 
\author{Jianguo~Bian$^{1}$}
\email{bianjg@mail.ihep.ac.cn}
\author{Jiahui~Wang$^2$}
\affiliation{\it
$^1$ Institute of High Energy Physics, Beijing 100049, China\\
$^2$ China Agricultural University,  Beijing 100083, China
}

\date{\today}

\begin{abstract}
 In the Standard Model of particle physics, the quarks and anti-quarks have fractional
 charge equal to $\pm1/3$ or $\pm2/3$ of the electron's charge. There 
 has been a large number of experiments searching  for fractional charge,
 isolatable, elementary  particles using a variety of methods, including
 $e^+e^-$ collisions  using dE/dx ionization energy loss measurements,
 but no evidence has been  found to confirm existence of  free fractional
 charge particles, which leads to the quark confinement theory. In
 this paper, a proposal to search for this kind particles is 
 presented, which is based on the conservation  law of four-momentum.
 Thanks to the CLEOc and BESIII detectors' large coverage, good particle identification, precision measurements
 of tracks' momenta  and their large recorded data samples, these features make 
 the scheme feasible in practice. The advantage of the scheme is 
 independent of any theoretical models and sensitive for 
 a small fraction of the quarks transitioning to the unconfinement  phase from
 the confinement phase.
 
\end{abstract}


\maketitle

 In the Standard Model (SM) of particle physics, the mesons comprise two of quarks and anti-quarks and the baryons comprise  three of quarks and anti-quarks,  
while the quarks and anti-quarks carry  fractional charge $\pm1/3e$ or $\pm2/3e$, $e$ being the magnitude of the electron charge$^{[1]}$.
The successes of SM in the fields of the spectroscopy and the high energy interaction of the hadrons have motivated a lot of
experimental searches for  fractional charge, isolatable, lepton-like particles. There has been a large number of experiments searching for free
fractional charge  particles using a variety of methods and no evidence up to date has been found to confirm their existence. 
There are three kinds of experiments used to search for fractional charge particles, which are based on the collision products of 
accelerator beams$^{[3-6]}$, cosmic rays$^{[7]}$, and bulk matters$^{[8]}$  respectively. No observation of fractional charge particles results
 in the confinement concept that the quarks are permanently confined within the hadrons.
The questions are whether there is a small fraction of quarks transitioning to the unconfinement phase from the confinement phase and whether there 
are new sensitive methods to search for them.
 
It is noted that dE/dx ionization energy loss measurements are used to identify charged particle species for $e^+e^-$$^{[3,4]}$ 
(or proton anti-proton$^{[5]}$ or heavy ions$^{[6]}$) collisions in accelerator based experiments. The energy loss  fluctuation is described by  
Landau's theory$^{[9]}$. One wonders if the theory  is still valid  for quarks, which take part in both the electroweak interaction and the strong 
interaction and is sensitive enough to distinguish the quarks from
the five stable particles such as the electron, muon ,pion, kaon and proton if there is a small fraction of  quarks which are unconfined. 
 However it is no doubt that the conservation law of momentum and energy (four-momentum)  applies to all kinds of  interactions,
which can be used to search for  fractional charge particles in combination with
the Lorentz force law for low energy accelerator experiments.  The prerequisite for this kind of search is that
a detector is required to possess large coverage, good particle identification, precise momentum resolution of charged tracks and energy 
resolution of neutral tracks  and collect large  data samples.  For instance, recently upgraded CLEOc$^{[10]}$ and BESIII$^{[11]}$ detectors can search for  fractional charge 
particles based on the laws. The center masses energy of the $e^+e^-$ collisions in CLEOc and BESIII cover the energy range of the charmonium production. Therefore
fractional charge particles in the mass range of 0 to 2 $GeV$ can be conceived to appear in the products of  the collisions if they are physical reality.

The momentum measurement of a track with charge $qe$  produced in 
an  $e^+e^-$ event is based its bending radius $R$ in the magnetic field $B$ along $z$ direction in the tracking chamber according 
to the Lorentz force law,
$$P_{xy}= qeB/R,$$
where $P_{xy}$ is the transverse component of the track momentum.
The longitudinal component is measured by considering 
 the polar angle $\theta$ of the track with respect to the $e^+$ beam direction, 
$$P_z=P_{xy}{\rm ctg}(\theta).$$
In other words, the radius  is a measured value while the momentum
is a derived value and dependent on the assumed charge $q$.
In the CELOc and BESIII experiments$^{[10,11]}$ as well as  others$^{[3-6]}$
$q$ is a priori  set to be 1 no matter how large charge it has.  
The momentum $\vec{P}_{\rm n}$ measured in this way, we call it the nominal momentum,  
  for the track equals to its real momentum $\vec{ P}_{\rm r}$ if $q$ is 1. If $q$ is not 1, its nominal  momentum 
changes by a factor of $1/q$ compared with its real momentum, i.e. $\vec{P}_{\rm n} = 1/q\vec{P}_{\rm r}$.

Hereafter $q$ is assumed to be less than 1 for simplicity. For events with at least one particle of charge larger than 1,
the similar analysis can be performed.
Assuming there is a pair of quarks $f$ and $\overline{f}$ with opposite charge $qe$ in the event such as $\pi^+\pi^- u\overline{u}$, $K^+K^- d\overline{d}$ and $p\overline{p}s\overline{s}$, which is
required to have even number of charged tracks. 
The total nominal  momentum and energy are derived as 
$$\vec{P}_{\rm n}{\rm (tot)}=\vec{P}_{\rm n}(f)+\vec{P}_{\rm n}(\overline{f})~+~{\rm others}$$
$$=1/q\vec{P}_{\rm r}(f)+1/q\vec{P}_{\rm r}(\overline{f})~+~{\rm others}~,$$
$$E_{\rm n}{\rm (tot)}=\sqrt{ P_{\rm n^2}(f)+m^2_f}+\sqrt{P_{\rm n^2}(\overline{f})+m^2_f}~+{\rm ~others}$$
$$=\sqrt{(1/qP_{\rm r})^2(f)+m^2_f}+\sqrt{(1/q P_{\rm r})^2(\overline{f})+m^2_f}~+{\rm ~others}.$$
If $\vec{P}_{\rm n}{\rm (tot)}\neq 0$, there are two cases. One is that some tracks in the event are undetected for they
are neutrinos or go beyond the coverage of the detector. Another is that $q$ is not 1. For the latter case,
 $q$ can be adjusted so that the real total momentum
$$\vec{P}_{\rm r}{\rm (tot)}=q\vec{P}_{\rm n}(f)+q\vec{P}_{\rm n}(\overline{f})~+{\rm ~others}=0.$$
Then the mass $m_f$ of the quark (anti-quark) can be calculated by requiring the real total energy
$$E_{\rm r}{\rm (tot)}=\sqrt{(qP_{\rm n})^2(f)+m^2_f}+\sqrt{(qP_{\rm n})^2(\overline{f})+m^2_f}~+{\rm ~others}=\sqrt{s},$$
there $\sqrt{s}$ is the mass center  energy of the $e^+e^-$ collisions.
In practical measurement and calculation, one does not know which pair of particles with opposite charge
have fractional charge. The momenta and mass of any one pair of particles with opposite charge can be adjusted so that
the event satisfies the conservation law of momentum and energy. It should be noted 
that $\vec{P}_{\rm r}{\rm (tot)}$ is a vector with three components, one cpmponent can be used to derive $q$, the two others can be used to suppress events from the first case because they can not satisfy the conservation law of 
three 
components simultaneously while events
with fractional charge particles can. 
The procedure is repeated for all events in the data sample and the parameter set of $(q,~m_f)$ 
is obtained. Then the parameter set is  plotted into a two dimension distribution. 
For the first case, $(q,~m_f)$ are a set of random numbers, but for the latter case,
they should concentrate at one or more points if there are one or more fractional charge particles in the data sample.
If fractional charge particles carry continuum masses, they should concentrate along one or more lines.

To suppress      the contribution to $(q,~m_f)$ from the first case further, the velocity $v_m$ of each of the pair of particles, which is  measured by the detector, can be compared with
the derived velocity $v_{\rm r}=qP_{\rm n}/\sqrt{m^2_f+(qP_{\rm n})^2}$. They should be consistent, i.e. $v_{\rm r}-v_{\rm m}=0$ if it is a  real fractional charge particle.
If the parameter set of $(q,~v_{\rm r} -v_{\rm m})$ is plotted
into two dimension distribution, the points corresponding to fractional charge particles should concentrate along $q$ axis. 

To search for  events containing a pair of quarks $f1$ and $\overline{f}2$ with opposite charge $qe$ and different masses $m_{f1}$ and $m_{\overline{f}2}$ such $d\overline{s}$,  the events are required to
have even number of charged tracks. The total momentum, energy and one of the velocities 
$$\vec{P}_{\rm r}{\rm (tot)}=q\vec{P}_{\rm n}(f1)+q\vec{P}_{\rm n}(\overline{f}2)~+{\rm ~others}~=0,$$
$$E_{\rm r}{\rm (tot)}=\sqrt{(qP_{\rm n})^2(f1)+m^2_{f1}}+\sqrt{(qP_{\rm n})^2(\overline{f}2)+m^2_{\overline{f}2}}~+{\rm ~others}=\sqrt{s},$$
$$v_{\rm r}(f1)=qP_{\rm n}/\sqrt{m^2_{f1}+(qP_{\rm n})^2(f1)}=v_{\rm m}(f1)$$
are used to derive $q$, $m_{f1}$ and $m_{\overline{f}_2}$. Another velocity
$v_{\rm r}(\overline{f}2)=qP_{\rm n}/\sqrt{m^2_{\overline{f}2}+(qP_{\rm n}/c)^2(\overline{f}2)}=v_{\rm m}(\overline{f}2)$ is used to suppress the contribution
 from the first case.

To search for  events containing a pair of quarks $f1$ and $f2$ with same sign  charge $qe$ and $(1-q)e$ and different masses 
$m_{f1}$ and $m_{f2}$ such as $u\overline{d}$ and $u\overline{s}$, the events
are required to have odd number of charged  tracks. The total momentum, energy and one of the velocities
$$\vec{P}_{\rm r}{\rm (tot)}=q\vec{P}_{\rm n}(f1)+(1-q)\vec{P}_{\rm n}(f2)~+{\rm ~others}=0,$$
$$E_{\rm r}{\rm (tot)}=\sqrt{(qP_{\rm n})^2(f1)+m^2_{f1}}+\sqrt{((1-q)P_{\rm n})^2(f2)+m^2_{f2}}~+{\rm ~others}=\sqrt{s},$$
$$v_{\rm r}(f1)=qP_{\rm n}/\sqrt{m^2_{f1}+(qP_{\rm n})^2(f1)}=v_{\rm m}(f1)$$
are used to derive $q$, $m_{f1}$ and $m_{f_2}$. Another velocity
$v_{\rm r}(f2)=(1-q)P_{\rm n}/\sqrt{m^2_{f2}+((1-q)P_{\rm n})^2(f2)}=v_{\rm m}(f2)$ is used to suppress the contribution from the first case.

To search for  a pair of  quarks $f$ and $\overline{f}$ with opposite charge $qe$ in events with only two charge tracks,
$$E_{\rm r}{\rm (tot)}=\sqrt{(qP_{\rm n})^2(f)+m^2_f}~+~\sqrt{(qP_{\rm n})^2(\overline{f})+m^2_f}=\sqrt{s},$$
$$m_f=qP_{\rm n}(f)\sqrt{1-v^2_{\rm m}(f)}/v_{\rm m}(f)$$
are used to derive $q$ and  $m_f$. Another velocity
$v_{\rm r}(\overline{f})=qP_{\rm n}/\sqrt{m^2_{f}+(qP_{\rm n}/c)^2(\overline{f})}=v_{\rm m}(\overline{f})$ is used to suppress the contribution
 from the first case.

Search for events containing  three free quarks or one free quark and one diquark  can be done
 in the similar way.  The two simplest examples are 
$\Delta^{++}\overline{u}\overline{u}\overline{u}~+~others~$ and 
$\Delta^{-}\overline{d}\overline{d}\overline{d}~+~others$ and their charge conjugation,
where $\Delta$ decays to $N\pi$.

To estimate the power of the technique above, it can be used to select fractional charge particles
from a Monte Carlo sample set,  which includes a signal sample  and a background sample. In this article,
200 events of $\psi(2S)\rightarrow d\overline{d}J/\psi,~J/\psi\rightarrow e^+e^-$ and
200 events of $\psi(2S)\rightarrow u\overline{u}J/\psi,~J/\psi\rightarrow e^+e^-$ 
through BESIII detector simulation$^{[11]}$ construct the signal sample. The background sample is 40 million events of $\psi(2S)$ inclusive decays,
 in which each decay branching ratio is from the particle data group (PDG). $d~(\overline{d})$ and $u~(\overline{u})$ carry 1/3 and 2/3 of 
the electron charge respectively. Their masses are assumed to be 0.005, 0.025, 0.045, 0.065 and 0.085 $GeV$. 

To pick up the signal events and suppress the background events, the events in the sample set are selected if they
 have four charge tracks with zero total charge. The scalar sum and the vector sum of the four tracks' 
momenta are required to larger than 3.8 $GeV$ and larger than 0.12 $GeV$ to remove the background events with 
normal four tracks. Then a pair of opposite charge tracks' momenta are adjusted by multiplying a factor of $q$ 
so that the event satisfy the momentum conservation. The events remain if the adjusted total momentum is less than 0.017 $GeV$, i.e.
$$P_{\rm r} {\rm (tot)}=\sqrt{\sum_{i=x,~y,~z}(qp_i(f)+qp_i(\overline{f})+p_i(e^+)+p_i(e^-))^2}
<0.017~GeV,$$
Where $f$ and $\overline{f}$ are supposed to be a pair of opposite fractional charge particles,
$ p(f),~p(\overline{f}),~p(e^+)$ and $p(e^-)$ are measured momenta for the four tracks.
If more than one pair of opposite charge tracks satisfy the requirement, the combination with minimal adjusted total momentum
 is chosen. The unadjusted pair is supposed to be a pair of $ e^+ e^- $ and is required to have 
invariant mass $\mid m_{e^+e^-}-3.097\mid <0.100$ $GeV$. The electron (positron) track is required to have zero hit in the muon detector. 

Instead to calculate the $f$ and  $\overline{f}$ masses,
the masses  squared are derived to be 
$m_f^2=q^2p^2(f)(1-v^2(f))/v^2(f),$
$m_{\overline{f}^2}
 =q^2p^2(\overline{f})(1-v^2(\overline{f}))/v^2(\overline{f}),$
where $v(f)$ and $v(\overline{f})$ are the measured velocities and often larger than 1 due to measurement errors.

Then the total energy
is required to be
$\mid E{\rm (tot)}-3.686\mid <0.15~GeV,$
where $$E{\rm (tot)}=\sqrt{(qp)^2(f)+m^2_f}+\sqrt{(qp)^2(\overline{f})~+~m^2_{\overline{f}}}~+~\sqrt{(p)^2(e^+)+m^2_{e^+}}~+~\sqrt{(p)^2(e^-)+m^2_{e^-}} .$$ 

The number of events  for the remaining events as well as  the measured charge $q$ and mass $m_f$ ($m_{\overline{f}}$) for the signal channels 
are listed in table 1 and table 2. The efficiency to select the first channel is larger than that of the second channel, because the $d$ and $\overline{d}'s$
bending radiuses in the detector are two time those of the $u$ and $\overline{u}$ if they
 carry the same real momentum and the $d$ and $\overline{d}$ arrive at the subdetector of the flight time easier.
The Fig. 1 shows the distribution of the charge $q_d$ and Fig. 2 shows the scatter of the mass  $m_d$ ($m_{\overline{d}}$) versus the charge $q_d$.
It can be seen from the two plots that the measured charge $q_d$ concentrates at 0.333 and  mass $m_d$ concentrates 0.010 $GeV$ for the channel $\psi(2S)\rightarrow d\overline{d}J/\psi$
with the input mass $m_d=0.005~ GeV$.
 22 events remain from the background sample, which are also plotted in Fig.1 and Fig. 2.  There is only one event in the $d$ charge window $\mid q_d -0.333\mid<0.040$
and 4 events in the $u$ charge window $\mid q_u-0.666\mid<0.040.$

\begin{table}
\begin{center}
Table 1{\hskip 0.4cm} The measured charge $q_d$, mass $m_d$ ($m_{\overline{d}}$)  $(GeV)$ and events for the channel $\psi(2S)\rightarrow d\overline{d}J/\psi$
{\vskip 0.2cm}
{\vskip 0.2cm}
\begin{tabular}{|c|c|c|c|c|c|}
\hline
 input mass $m_d$          &0.005                & 0.025                &0.045                & 0.065             & 0.085    \\
\hline
 output mass $m_d^2$     &$0.010^2\pm0.026^2$  & $0.022^2\pm0.026^2$  &$0.037^2\pm0.031^2$  &$0.053^2\pm0.026^2$    &$0.078^2\pm0.029^2$              \\
\hline
 charge  $q_d$                 &$0.339\pm0.003$      & $0.333\pm0.002$      &$0.333\pm0.003$      &$0.329\pm0.003$    &$0.323\pm0.002$           \\
\hline
 events              &36                   & 45                   & 38                  & 32                 & 38            \\
\hline
\end{tabular}
\end{center}
\end{table}

\par
{\vskip 0.2cm}

\begin{table}
\begin{center}
Table 2{\hskip 0.4cm} The measured charge $q_u$, mass $m_u$ ($m_{\overline{u}}$) $(GeV)$ and events for the channel $\psi(2S)\rightarrow u\overline{u}J/\psi$
{\vskip 0.2cm}
{\vskip 0.2cm}
\begin{tabular}{|c|c|c|c|c|c|}
\hline
 input mass $m_u$            & 0.005               & 0.025                & 0.045               &0.065                  & 0.085    \\
\hline
 output mass $m_u^2$       &$0.016^2\pm0.031^2$  &$0.033^2\pm0.029^2$   &$0.044^2\pm0.031^2$  &$0.064^2\pm0.029^2$    &$0.072^2\pm0.024^2$     \\
\hline
charge  $q_u$                   &$0.635\pm0.006$      &$0.662\pm0.005$       &$0.660\pm0.004$      &$0.654\pm0.009$         &$0.659\pm0.004$        \\
\hline
events                 & 28                  &24                    &33                   & 32                     &27      \\
\hline
\end{tabular}
\end{center}
\end{table}
\par
{\vskip 0.2cm}

\begin{figure}[hbtp]
  \begin{center}
    \resizebox{10cm}{6cm}{\includegraphics{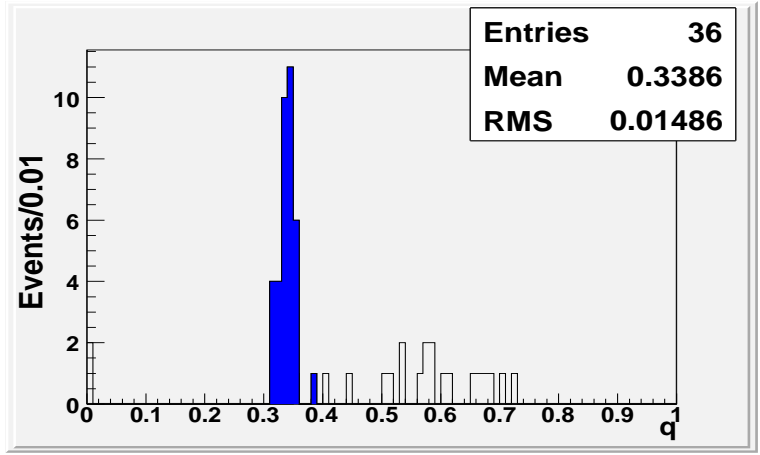}}
    \caption{The distribution of the measured charge $q_d$ for $\psi(2S)\rightarrow d\overline{d}J/\psi,~J/\psi\rightarrow e^+e^-$ with the input $m_d=0.005~GeV.$
The shaded area is the signal events. The others are background events. The statistics is only for the signal events.
}
    \label{fig:ex1}
  \end{center}
\end{figure}

\begin{figure}[hbtp]
  \begin{center}
    \resizebox{10cm}{6cm}{\includegraphics{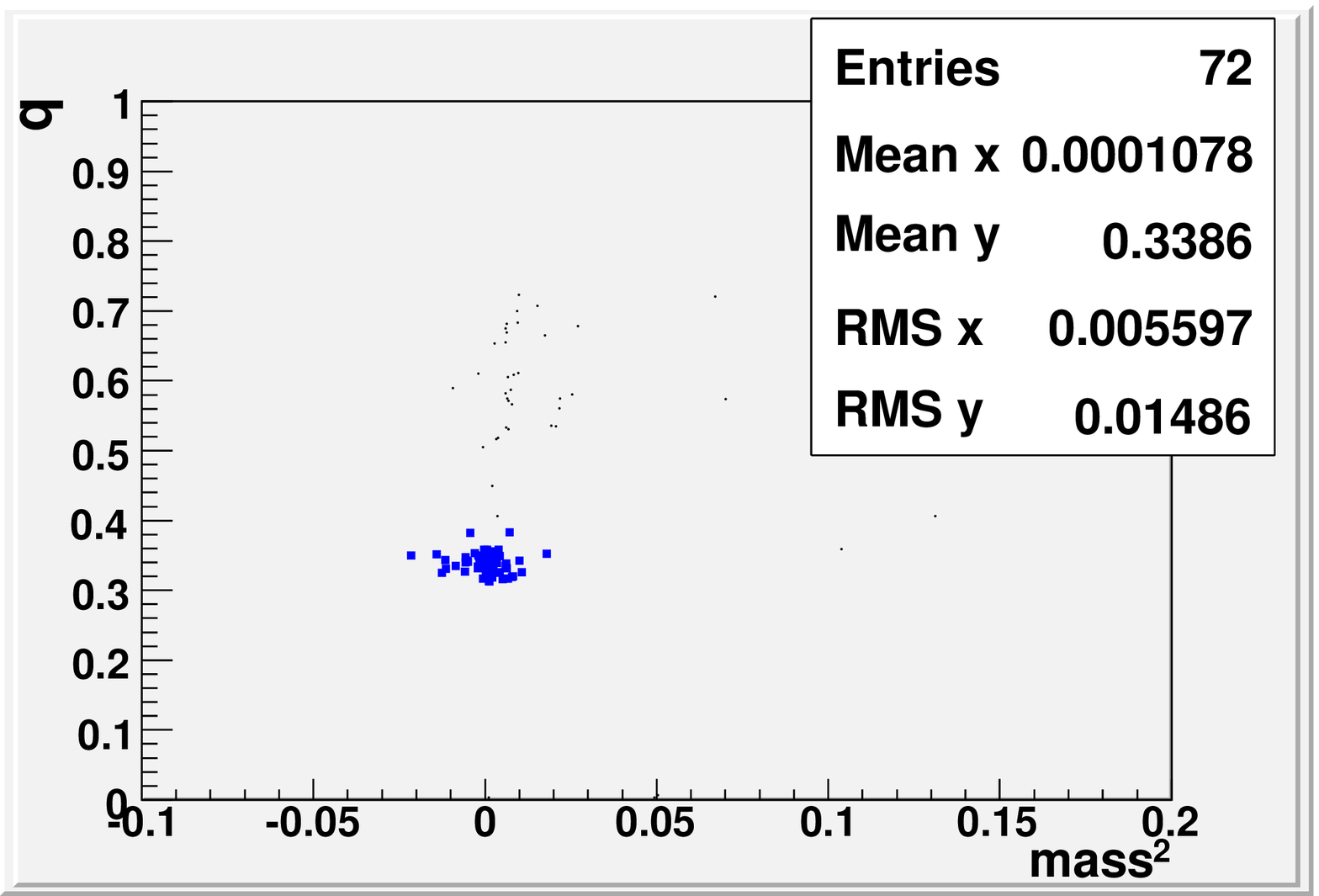}}
    \caption{The scatter of the measured $m_d^2$ versus the charge $q_d$ for $\psi(2S)\rightarrow d\overline{d}J/\psi,~J/\psi\rightarrow e^+e^-$ with the input $m_d=0.005~GeV.$
The squares are the signal events. The others are background events. The statistics is only for the signal events.
 }
    \label{fig:ex2}
  \end{center}
\end{figure}

In summary,
 an inclusive sample can be divided into  exclusive channels $n_1 \pi s~+~n_2 Ks~+~n_3 p s~+~n_4 e^+e^-s~+~n_5 \mu^+\mu^- s~+~n_6 \gamma s~+~ n_7 fs,$
where $f$ denotes a fractional charge particle. 
For a  fractional charge track, its nominal momentum, derived by assuming its charge equal to 1 when it is reconstructed,
increases by a factor $1/q$ compared with its real momentum. Then the nominal total momentum and energy for the event 
containing  this kind of tracks will not satisfy the conservation law, which can help find 
fractional charge particles described by the two parameters $(q,~m_f)$ by analyzing each exclusive channel. 
Between the two parameters,
 $q$ is more sensitive than $m_f$. Whether the distribution of $m_f$ concentrates      at one or more points very well depends on
the momentum  resolution for charged tracks and the energy resolution for neutral tracks.
The effect of $\mid 1/q-1\mid $ on the momentum for charged tracks  is much larger than the resolutions. Let $q=2/3$, 
then $\mid1/q-1\mid$ is $50\%$, while the momentum resolution is $0.5\%$ for charged tracks and $2.5\%$ for neutral tracks for BESIII$^{[11]}$.
 The larger $\mid 1/q-1\mid$ is, the more sensitive the method is to 
search for the  fractional charge particles, especially for the type of $f\overline{f}$.  Another advantage of the method is independence of any theory
 models.  Even if fractional charge particles carry continuum masses, the method can be used to search for them, while
the dE/dx ionization energy loss method is not adequate.

\end{document}